\documentclass[journal=nalefd,manuscript=letter 
]{achemso}

\usepackage[T1]{fontenc} 
\usepackage{graphicx}  
\usepackage{bm}        
\usepackage{amssymb}   
\usepackage{amsmath}
\usepackage[colorlinks=true, pdfstartview=FitV, linkcolor=blue, citecolor=blue, urlcolor=blue]{hyperref} 
\usepackage{textcomp}
\usepackage{upgreek}

\newcommand*{\Hydrogen}{{$^1$H}}

\newcommand*{\ee}[1]{\cdot10^{#1}}
\newcommand*{\unit}[1]{\,\mathrm{#1}}
\newcommand*{\yn}{\gamma_\mathrm{n}}
\newcommand*{\braket}[1]{\ensuremath{\langle#1\rangle}}
\newcommand*{\Bonemax}{B_1}
\newcommand*{\Bonemod}{B_{1,\mathrm{mod}}}
\newcommand*{\Df}{\Delta f}
\newcommand*{\dxB}{\partial_x|\vecB|}
\newcommand*{\Fspin}{F_\mathrm{spin}}
\newcommand*{\Fspinmax}{F_\mathrm{spin,max}}
\newcommand*{\Fnoise}{F_\mathrm{noise}}
\newcommand*{\fc}{f_\mathrm{c}}
\newcommand*{\fFM}{f_\mathrm{FM}}
\newcommand*{\fL}{f_\mathrm{L}}
\newcommand*{\frf}{f_\mathrm{rf}}
\newcommand*{\frfmod}{f_\mathrm{rf,mod}}
\newcommand*{\fidelity}{\mathcal{F}}
\newcommand*{\Gx}{G_x}
\newcommand*{\HS}{\mathrm{HSn}}
\newcommand*{\sigxslope}{\sigma_{x,\mathrm{slope}}}
\newcommand*{\sigxo}{\sigma_{x_0}}
\newcommand*{\Tc}{T_\mathrm{c}}
\newcommand*{\vecB}{\mathbf{B}}
\newcommand*{\vecG}{\mathbf{G}}
\newcommand*{\vecr}{\mathbf{r}}
\newcommand*{\vecrd}{\mathbf{r}'}
\newcommand*{\vecnabla}{\bm{\nabla}}
\newcommand*{\T}{\mathcal{T}}

\DeclareMathOperator{\sech}{sech}

\author{U. Grob}
\altaffiliation{These authors contributed equally to this work.}
\author{M. D. Krass}
\altaffiliation{These authors contributed equally to this work.}
\author{M. H\'eritier}
\author{R. Pachlatko}
\author{J. Rhensius}
\author{J. Ko\v{s}ata}
\author{B. A. Moores}
\affiliation{Department of Physics, ETH Zurich, Otto Stern Weg 1, 8093 Zurich, Switzerland.}
\alsoaffiliation{Present address: JILA, National Institute of Standards and Technology and the University of Colorado, Boulder, Colorado 80309, USA and Department of Physics, University of Colorado, Boulder, Colorado 80309, USA.}
\author{H. Takahashi}
\affiliation{Department of Physics, ETH Zurich, Otto Stern Weg 1, 8093 Zurich, Switzerland.}
\alsoaffiliation{Present address: JEOL RESONANCE Inc., Musashino, Akishima, Tokyo 196-8558, Japan.}
\author{A. Eichler}
\author{C. L. Degen}
\email{degenc@ethz.ch}
\affiliation{Department of Physics, ETH Zurich, Otto Stern Weg 1, 8093 Zurich, Switzerland.}

\title{Magnetic resonance force microscopy with a one-dimensional resolution of 0.9 nanometers}

\abbreviations{MRFM,NMR,AFM,SPM,MRI}

\begin{document}


\begin{abstract}
Magnetic resonance force microscopy (MRFM) is a scanning probe technique capable of detecting MRI signals from nanoscale sample volumes, providing a paradigm-changing potential for structural biology and medical research.  Thus far, however, experiments have not reached sufficient spatial resolution for retrieving meaningful structural information from samples.  In this work, we report MRFM imaging scans demonstrating a resolution of $0.9\unit{nm}$ and a localization precision of $0.6\unit{nm}$ in one dimension.
Our progress is enabled by an improved spin excitation protocol furnishing us with sharp spatial control on the MRFM imaging slice, combined with overall advances in instrument stability.  From a modeling of the slice function, we expect that our arrangement supports spatial resolutions down to $0.3\unit{nm}$ given sufficient signal-to-noise ratio.  Our experiment demonstrates the feasibility of sub-nanometer MRI and realizes an important milestone towards the three-dimensional imaging of macromolecular structures.
\end{abstract}

\section{Introduction}
The goal of nanoscale magnetic resonance imaging (``nano MRI'') is the three-dimensional visualization of nuclear spin densities inside materials with near-atomic spatial resolution and on length scales of up to a few $100\unit{nm}$.  Such images are expected to provide fundamental insight into the structure and composition of matter, especially in the field of structural molecular biology. For example, nano MRI images could serve as templates for modeling the global arrangement of large protein assemblies. If sub-nanometer resolution can be realized, nano MRI could even allow for a direct imaging of tertiary or secondary protein structure, and ultimately, the atomic arrangement.  Important advantages of MRI compared to other structural imaging techniques, such as electron tomography, are its high chemical selectivity and the fact that single copies of molecules can be imaged in a non-destructive manner.

One promising candidate for nano MRI is magnetic resonance force microscopy (MRFM)~\cite{sidles91,sidles95,degen09,poggio10,nichol13}. The method employs an ultrasensitive nanomechanical transducer to detect the interaction between nuclear spins in the sample and a nanoscale magnetic tip by means of a magnetic force.  Thanks to major advances in mechanical transduction~\cite{mamin01,moser13,tao14,rossi17,delepinay17,heritier18} and magnetic gradient generation~\cite{mamin12apl,nichol12,longenecker12,mamin12apl,tao16,wagenaar17}, researchers have in recent years greatly improved the sensitivity of MRFM.  Latest imaging experiments reported sensitivities of order $50-100$ nuclear moments, corresponding to voxel sizes between $(3.5\unit{nm})^3 - (5.5\unit{nm})^3$ for statistically polarized protons in organic material \cite{mamin09,rose18}.  In principle, MRFM even offers the sensitivity required to detect single proton magnetic moments \cite{tao16}, but it is unclear at present whether this sensitivity can be extended to the context of three-dimensional MRI.  Unlike other nano MRI techniques, like nitrogen-vacancy-center NMR \cite{mamin13,staudacher13}, MRFM is capable of handling larger objects ($>20\unit{nm}$) by adjusting the sizes of the mechanical sensor and the magnetic tip.

In order to provide meaningful information about the macromolecular arrangement of protein complexes, the spatial resolution must be of order $\sim 1\unit{nm}$.  Despite providing sufficient detection sensitivity, recent MRFM scans have fallen short of this goal. Imaging experiments on single tobacco mosaic virus particles \cite{degen09} have reported a best-effort resolution of $4\unit{nm}$ in one dimension, limited by a combination of scan step size, available magnetic gradient, thermomechanical force noise, and instrument stability.  Other experiments have shown feature sizes of order $5-10\unit{nm}$ \cite{mamin09,moores15}.  Recently, Rose {\it et al.} \cite{rose18} have reported a nominal resolution of $\sim 2\unit{nm}$ for a polystyrene-coated nanowire using a novel Fourier encoding method.

In this paper, we demonstrate MRFM scans with a one-dimensional resolution of $0.9\unit{nm}\pm 0.2\unit{nm}$ and a localization precision of $0.6\unit{nm}\pm 0.1\unit{nm}$.  This progress is enabled through: (i) an improved spin inversion protocol providing a sharply defined imaging slice, (ii) the use of a state-of-the-art magnetic gradient exceeding $10^6\unit{T/m}$, (iii) the use of a suitable nanorod sample geometry, and (iv) a high instrument stability with drifts of less than $2\unit{nm}$ over 24h. As a result of our efforts, our experimental resolution is limited solely by the sensor noise (which we have not improved within this work) and not by other factors.  Our demonstration is achieved with a top-down fabricated cantilever that is suitable for loading large molecules and amenable to modular sample preparation techniques.  With this result, we take a critical step towards three-dimensional imaging of biological samples on the $\sim 1\unit{nm}$ lengthscale.

\section{Experimental Setup}

\begin{figure*}
\includegraphics[width=\columnwidth]{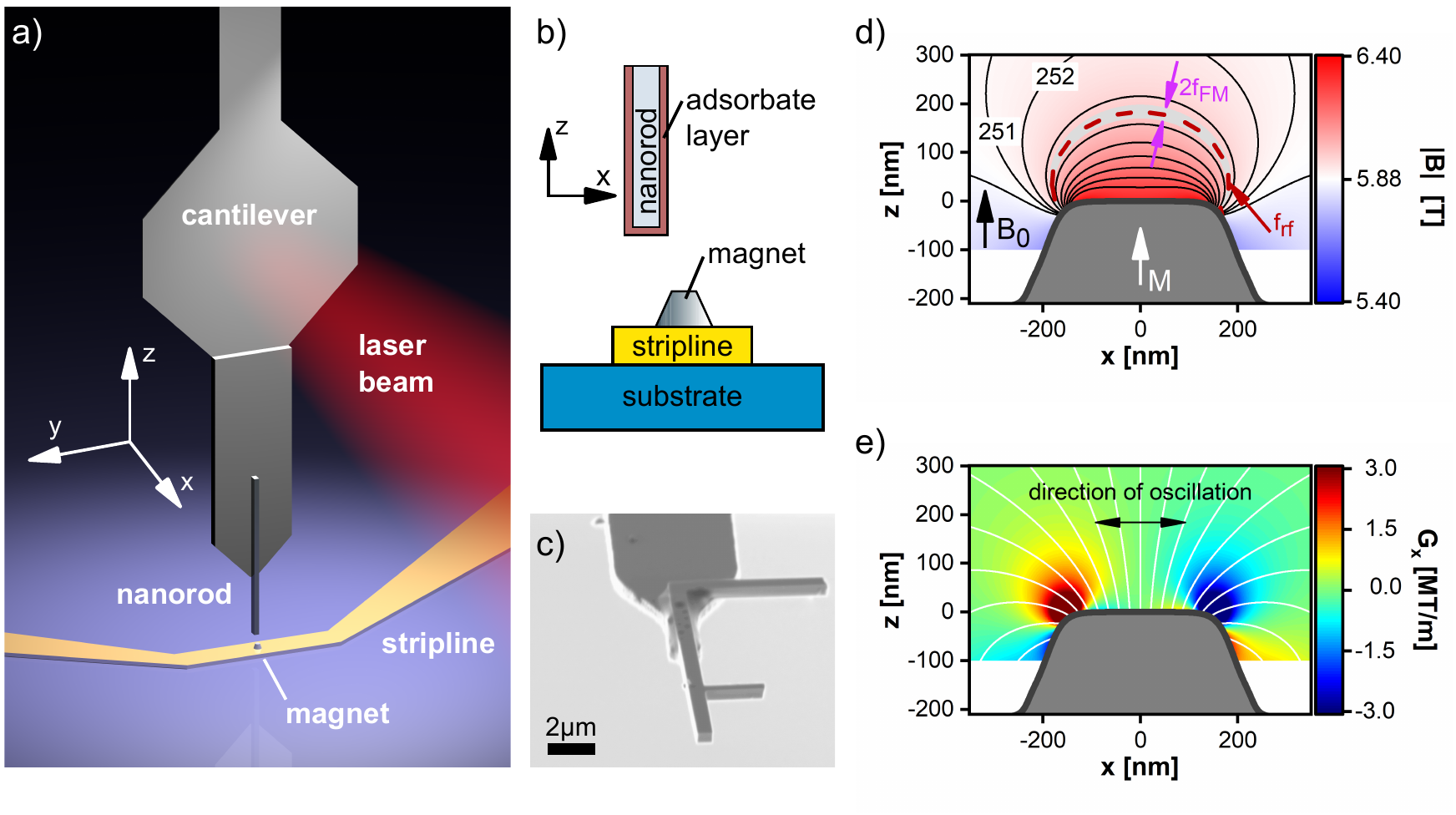}
\caption{\label{fig:figure1}
(a) Configuration of the MRFM experiment.
(b) Sketch of nanorod apex, $^1$H-rich adsorbate layer, magnetic tip and stripline in cross section.
(c) Scanning electron micrograph of a silicon nanorod and cantilever, similar to the device used in experiments.
(d) Calculated magnetic field map $|\vecB(x,z)|$ near the FeCo nanomagnet \cite{poggio07,moores15}.  An external bias field $B_0=5.88\unit{T}$ is applied along $z$ to polarize the nanomagnet and to stabilize the quantization axis of the nuclear spins.  Contours indicate the \Hydrogen{} NMR frequency in the given field.  The shaded region represents a resonant slice with rf pulse frequency $\frf = 255\unit{MHz}$ and FM modulation depth $\fFM = 500\unit{kHz}$.  The field map is calculated from a combination of AFM topography and MRFM calibration scans (see SI).
(e) Magnetic gradient $\Gx \equiv \dxB$ calculated from the field map shown in panel d.
}
\end{figure*}

A schematic of our MRFM setup is shown in Fig.~\ref{fig:figure1}a.  The sample, in our case a silicon nanorod with a thin film of $^1$H-containing adsorbate molecules (see Fig.~\ref{fig:figure1}b,c), is affixed to the end of the cantilever force transducer and positioned above a nanomagnet.  The nanomagnet produces a highly localized magnetic field $B=|\vecB(\vecr)|$ and an associated strong magnetic gradient $\vecG=\vecnabla B$ (see Fig.~\ref{fig:figure1}d,e). The gradient generates an attractive or repulsive force on the nuclear spins in the sample. To measure the magnetic force, nuclear spins are periodically inverted by adiabatic radio-frequency (rf) pulses to drive the transducer at its kHz mechanical resonance. The output signal is provided by the oscillation amplitude of the cantilever, which is proportional to the spin force and which we detect by an optical interferometer.

As in conventional MRI, image generation takes advantage of a spatially localized excitation of nuclear spins inside the sample volume.  Radio-frequency pulses act selectively on spins whose Larmor frequencies $\fL=\yn B$ are within the excitation bandwidth of the rf pulse frequency $\frf\pm\fFM$.  Here, $\yn$ is the nuclear gyromagnetic ratio ($\yn = 42.58\unit{MHz/T}$ for protons), $\frf$ is the rf carrier frequency and $\fFM$ is the modulation depth of rf pulses.  Due to the locally inhomogeneous field, only spins within a thin spatial slice are inverted and generate a driving force on the mechanical sensor (see Fig.~\ref{fig:figure1}d).  Specifically, the force signal is given by the convolution of the slice with the sample's three-dimensional spin density $\rho(\vecr)$ \cite{chao04,degen09,poggio10},
\begin{align}
\Fspin^2 = \mu^2 \int_{\mathbb{R}^3} \mathrm{d}\vecrd \rho(\vecr-\vecrd) G_x^2(\vecrd) \xi(\vecrd) \ .
\end{align}
where $\mu = h\yn/2$ is the nuclear magnetic moment, $h=6.63\ee{-34}\unit{J/Hz}$ is Planck's constant, $G_x$ is the $x$-component of the magnetic gradient, and $\xi(\vecrd)$ contains the shape of the spatial slice.  We remind that because we detect statistical rather than thermal spin polarization \cite{degen07,herzog14}, our signal is proportional to the variance $\Fspin^2$ of the spin force, rather than the magnitude.  The slice function $\xi(\vecrd) \equiv \xi(\fL-\frf) \equiv \xi(\Df)$ depends on the detuning $\Df = \fL-\frf$ between the (location-dependent) Larmor frequency of a nuclear spin, $\fL=\yn B(\vecrd)$, and the rf excitation frequency $\frf$.  We control $\xi(\Df)$ by tuning shape and parameters of the adiabatic rf pulse modulation.  In this work, we are particularly interested in the frequency selectivity of pulses, since this determines the sharpness of the slice edges.  At the same time, the rf excitation must be robust against variations in pulse amplitude, because hundreds of coherent spin reversals are required to build up a detectable oscillation of the mechanical sensor.

\begin{figure}
\includegraphics[width=0.48\columnwidth]{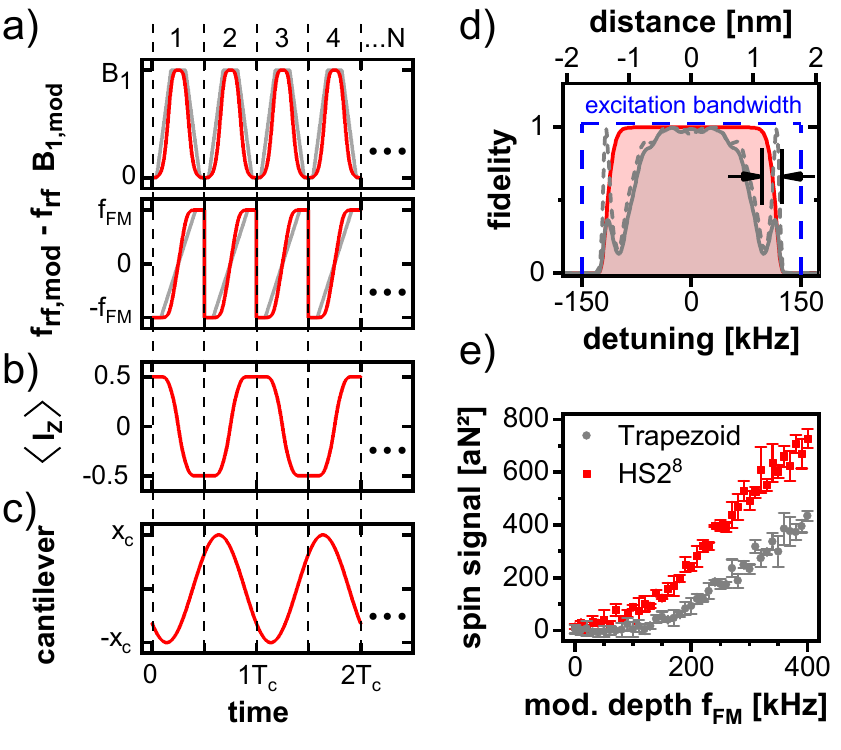}
\caption{\label{fig:figure2}
Signal encoding and generation by hyperbolic secant ($\HS$) spin reversals.
(a) Amplitude $\Bonemod(t)$ and frequency $\frfmod(t)$ of adiabatic rf pulses.  The $\HS$ modulation is shown in red and conventional trapezoidal modulation is shown in gray for comparison.  Parameters are $n= 2$ and $\beta = 8$.  $\Tc=1/\fc$ represents one cantilever oscillation period (here $\Tc\sim 300\unit{\text{\textmu}s}$).
(b) Simulated nuclear spin polarization $\langle\hat I_z(t)\rangle$ in response to the rf pulses in panel a.
(c) Cantilever oscillation amplitude in response to the driving force of panel b.
(d) Simulated fidelity $\fidelity(\Df)$ of spin reversals (Eq. \ref{eq:fidelity}) as a function of nuclear spin detuning $\Df = \fL - \frf$.  The excitation bandwidth is $2\fFM = 300\unit{kHz}$ (blue dashed line).  Slice profiles are shown for $\HS$ modulation with $\Bonemax=5.3\unit{mT}$ (red solid line), as well as for trapezoidal modulation with $\Bonemax=2.5\unit{mT}$ (gray solid line) and $\Bonemax=2.0\unit{mT}$ (gray dashed line).
The arrows indicate the sharp slice edge of $27\unit{kHz}$.  The top scale indicates the spatial slice width in a gradient of $|\vecG|=2\ee{6}\unit{T/m}$.
(e) Experimental MRFM signal as a function of the modulation depth $\fFM$ for $\HS$ (red) and trapezoidal (gray) modulation.  The rf carrier frequency is $\frf = 254\unit{MHz}$ and the rf amplitude is $\Bonemax = 5.3\unit{mT}$.
}
\end{figure}

In a first part of this study, we have explored several adiabatic rf modulation schemes for their ability to robustly invert spins in a narrow, well-defined frequency bandwidth. We find experimentally and through simulations that hyperbolic secant ($\HS$) pulses~\cite{silver85,tannus96,tomka13} are well-suited for this task. Other schemes that we explored include $\mathrm{Gauss}$ \cite{tannus96}, $\mathrm{WURSTn}$ \cite{Kupce95}, and $\mathrm{Sin/Cos}$ \cite{bendall86} modulation.  $\HS$ pulses involve both amplitude (AM) and frequency (FM) modulation of the rf field, 
\begin{align}
\Bonemod(\T) &= \Bonemax\, \sech(\beta\T^n) \\
\frfmod(\T) &= \frf + \fFM\,\left[c(n,\beta)\, \left(\int_{-1}^{\T}  \sech^2(\beta\T'^n)\, \mathrm{d}\T'\right)-1\right]
\end{align}
where $\Bonemax$ is the in-plane component of the rf field amplitude, $\lvert \T \rvert = \lvert \tfrac{4t}{T_c} \rvert \leq 1$ is a normalized time running over half a cantilever oscillation period, and $c(n,\beta)$ is a unit-less factor that normalizes the integral for a symmetrical modulation around $\frf$.  The parameters $n$ and $\beta$ are integers that control the truncation and steepness of the frequency modulation. The $\HS$ profile for $n=2$ and $\beta=8$ is shown in Fig.~\ref{fig:figure2}a.

We calculate slice functions $\xi(\Df)$ for the $\HS$ and other modulation schemes by simulating the spin reversal and computing the Fourier coefficient at the mechanical resonance $\fc$.  Fig.~\ref{fig:figure2}b shows the expectation value $\braket{\hat{I}_z(t)}$ of the single spin operator $\hat I_z$ during the reversal.  $\braket{\hat{I}_z(t)}$  is calculated using density matrices under a piece-wise constant Hamiltonian (see SI).  The first real Fourier coefficient is
\begin{align}
a_1 = \frac{4}{\Tc} \int_0^{\Tc/2} 2\;\braket{\hat{I}_z(t)}\cos(2\uppi\fc t)\ \mathrm{d}t\ .
\end{align}
The cantilever motion follows the periodic force generated by $\braket{\hat I_z(t)}$ as shown in Fig.~\ref{fig:figure2}c.  Because of order $N=2\tau/\Tc\sim 10^2-10^3$ reversals are needed to ring up the mechanical sensor, where $\tau$ is the resonator's time constant, spin inversions must be highly efficient with no loss of magnetization over many hundreds of cycles $N$.  We account for the inversion efficiency through the fidelity
\begin{align}\label{eq:fidelity}
\fidelity = \left[\braket{\hat{I}_z(\Tc/2)} - \braket{\hat{I}_z(0)} \right]^N ,
\end{align}
where $\braket{\hat{I}_z(0)}$ and $\braket{\hat{I}_z(\Tc/2)}$ are the expectation values of $\hat I_z$ at the beginning and the end of the inversion pulse, respectively.  For the $\HS$ modulation shown in Fig.~\ref{fig:figure2}a, all spins in the slice flip at approximately the same instance in time, therefore $a_1 \approx 4/\uppi$.
The slice function is then given by
\begin{align}
\xi(\Df) = a_1(\Df) \fidelity(\Df) \approx \frac{4}{\uppi} \fidelity(\Df)
\end{align}

In Fig.~\ref{fig:figure2}d we plot the fidelity $\fidelity(\Df)$ of spin inversions for a sequence of $N=140$ $\HS$ pulses as a function of the detuning from the slice center frequency, for a modulation depth of $\fFM = 150\unit{kHz}$.  This number of pulses $N$ approximately corresponds to the number of reversals within the $\tau \approx 20\unit{ms}$ ring-up time of the feedback-damped cantilever. The figure shows that $\HS$ pulses are clearly very effective at inverting spins.  Importantly, the frequency slice is sharply defined with only $\sim 30\unit{kHz}$ between complete inversion ($\fidelity = 1$) and no signal ($\fidelity=0$).  By comparison, common trapezoidal pulses with a linear frequency ramp~\cite{degen09,mamin09,moores15} produce an ill-defined slice profile due to the sudden turn-on of the rf amplitude (gray curves).  Further advantages of the $\HS$ modulation are its robustness in the presence of $B_1$ variations (see SI) and nuclear spin-spin interactions \cite{tomka13}.

\section{Results and Discussion}

\begin{figure}
\includegraphics[width=0.48\columnwidth]{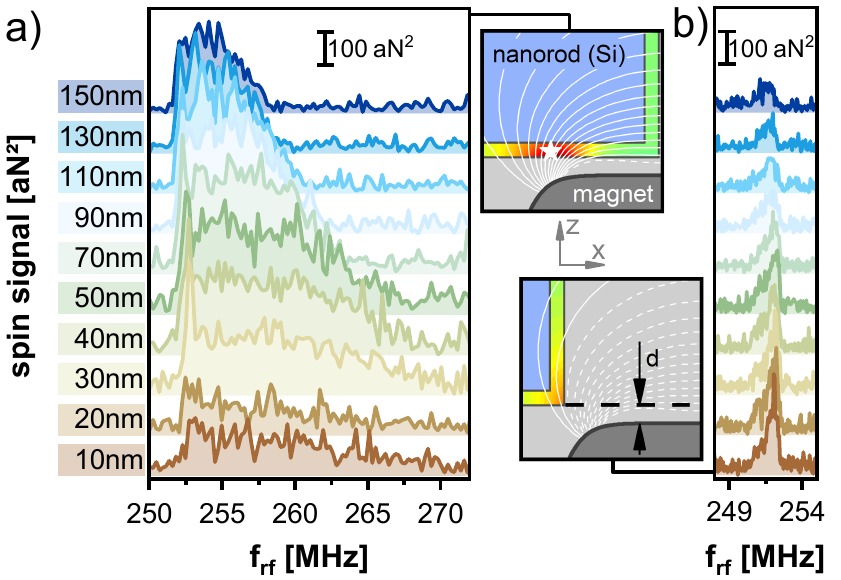}
\caption{\label{fig:figure3}
Series of \Hydrogen{} NMR spectra as a function of vertical approach distance $d=10-150\unit{nm}$. The schematics show the configuration of nanorod and nanomagnet; the adsorbate layer is color-coded with the gradient $\Gx$ (scale of Fig.~\ref{fig:figure1}e). 
(a) Spectra taken with the tip positioned over the edge of the nanomagnet ($x=-200\unit{nm}$, where $x$ is the center-to-center distance between cantilever and nanomagnet). The white star marks the position of highest gradient of $\sim 6\ee{6}\unit{T/m}$.
(b) Spectra taken with the tip positioned in front of the nanomagnet ($x=-400\unit{nm}$).
Modulation depth is $\fFM = 300\unit{kHz}$ in panel a and $\fFM = 150\unit{kHz}$ in panel b.
The bias field is $5.88\unit{T}$.
}
\end{figure}

We experimentally demonstrate high-resolution \Hydrogen{} NMR imaging using the $\sim 1\unit{nm-thick}$ adsorbate film \cite{degen09,mamin09,loretz14apl} on a silicon nanorod \cite{overweg15}.  The nanorod has an edge length of $\sim 500\unit{nm}$ (see Fig.~\ref{fig:figure1}b) and is fabricated using standard silicon lithography (see SI).  Our choice of test sample has two motivations:  first, the nanorod has a well-defined geometry and the natural adsorbate layer provides a strong \Hydrogen{} NMR signal.  Second, the nanorod can be batch produced and can undergo water dipping and shock-freezing.  This capability is important for preparing biological samples in future experiments.  For the present study, no sample is loaded onto the nanorod and the natural adsorbate film provides the only \Hydrogen{} NMR signal. The nanorod is glued to the end of an ultrasensitive silicon cantilever that is in turn mounted in the custom MRFM apparatus (Fig.~\ref{fig:figure1}b).  The cantilever used in this study has, after loading, a natural frequency of $\fc \sim 3.5\unit{kHz}$, a quality factor of $Q \sim 28,000$, and a spring constant of $k=82\unit{\text{\textmu}N/m}$ (see SI). The microscope is operated in high vacuum at the bottom of a cryogen-cooled helium cryostat ($T \sim 4.7\unit{K}$).  Further details on the apparatus are given in Ref. \citenum{moores15}.

In a first set of experiments, we examine the resonant slice profile and optimize the parameters of the $\HS$ protocol. In Fig.~\ref{fig:figure2}e, we show the signal magnitude $\Fspin^2$ as a function of modulation depth $\fFM$. We observe a roughly linear scaling of $\Fspin^2$ with $\fFM$, as expected, because the resonant slice volume, and hence the signal, are proportional to $\fFM$. We also find that modulation depths as small as $\fFM = 75\unit{kHz}$ still lead to a detectable signal. By using the gradient $G_x = 2.3\ee{6}\unit{T/m}$ extracted from the tip model in Fig.~\ref{fig:figure1}d and the simulated slice shape for the experimental settings, we find that the total slice width is only $0.7\unit{nm}$.\footnote{Note that the experimental settings were different from the ones shown in Fig.~\ref{fig:figure2}d; see SI for details.}  We have also measured the corresponding signal for a conventional trapezoidal pulse modulation; here, no signal is detectable for $\fFM < 150\unit{kHz}$ and detection is highly susceptible to $B_1$ miscalibration (see Fig.~\ref{fig:figure2}e).  We further examined $\HS$ pulses for a range of $n$, $\beta$ parameters; these results are provided as SI.  Figure~\ref{fig:figure2}e confirms that $\HS$ pulse modulation is well-suited for precise, high-resolution imaging.

We next record a series of \Hydrogen{} NMR spectra as a function of the vertical approach distance $d$.  These spectra serve to determine the optimum position and rf frequency $\frf$ for high-resolution imaging scans.  A first series of spectra, shown in Fig.~\ref{fig:figure3}a, is taken with the cantilever centered over the magnet's edge.  In this position, spectral peaks are broad, because a large number of slices penetrates the sample volume.  Peak widths exceeding $15\unit{MHz}$ are found for close approach distances ($d<40\unit{nm}$), corresponding to tip fields in excess of $450\unit{mT}$ (see SI).
In all spectra, the low-frequency ends ($\fL \approx 252\unit{MHz}$) contain signal from spins that experience little tip field, while the high-frequency ends reflect locations over the magnet where the tip field is high.

Figure~\ref{fig:figure3}b shows a second series of NMR spectra recorded with the cantilever positioned in front of the magnet. In this configuration, the sample surface and slice edge are oriented nearly tangentially (see schematic in Fig.~\ref{fig:figure3}b) and the spectra become narrow.  This tangential configuration is therefore well-suited for demonstrating high-resolution imaging scans along the $x$-direction, because a large portion of the sidewall \Hydrogen{} adsorbate layer can be moved into the imaging slice within a few nm.  This results in a large signal change over a short distance.

\begin{figure}
\includegraphics[width=0.48\columnwidth]{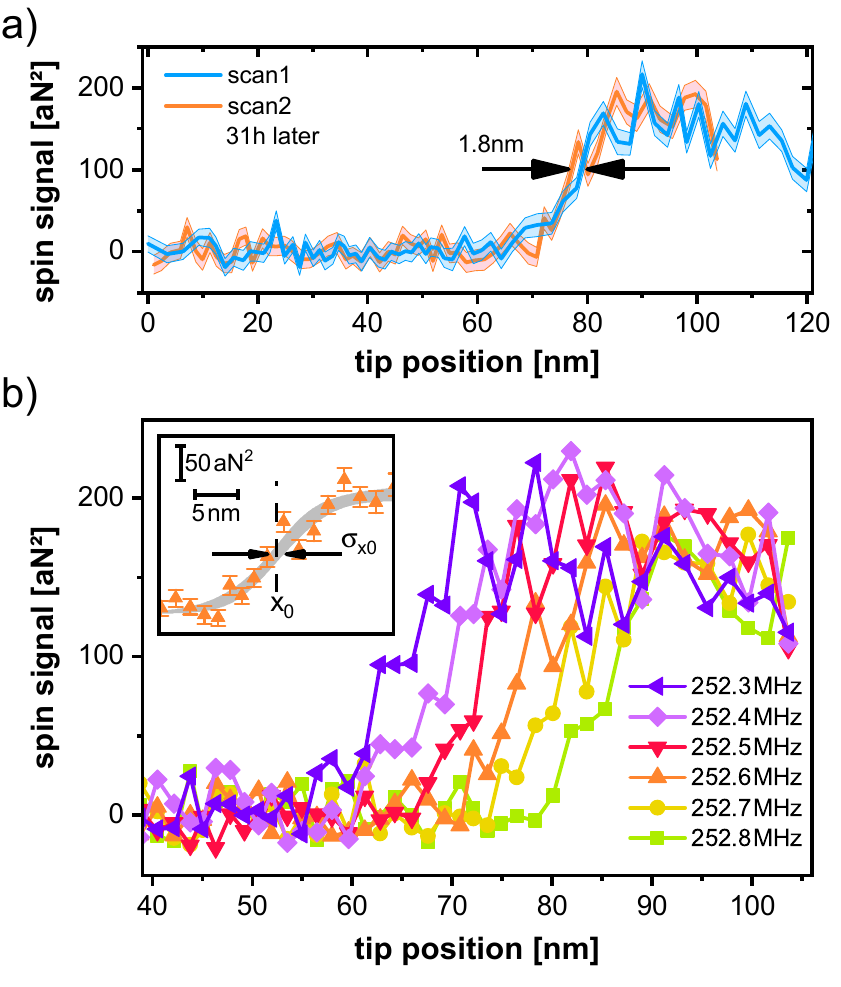}
\caption{\label{fig:figure4}
Lateral imaging scans showing $0.9\unit{nm}$ spatial resolution.
(a) Two consecutive line scans taken within 31 hours.  Averaging time is $4\unit{min}$ per point.
(b) Series of $x$ scans for excitation frequencies $\frf=252.3-252.8\unit{MHz}$.
The sharp signal rise around $x=20-50\unit{nm}$ indicates the position where the nanowire \Hydrogen{} layer enters the slice.
The scan height is $z=30\unit{nm}$. The inset shows the $90\unit{\%}$ confidence interval of the fit to Eq.~\ref{eq:fitfunc} for $\frf = 252.6\unit{MHz}$.
}
\end{figure}

Fig.~\ref{fig:figure4}a shows two lateral scans in the tangential nanorod configuration at $\fc=252.6\unit{MHz}$.  The nominal step size for these high-resolution scans is $1.6\unit{nm}$.  We calibrate the tip $x$ position by a separate line scan over the known stripline topography, and correct the nominal scanner position for static cantilever deflections caused by electrostatic tip-magnet interactions (see SI).  The sudden onset of signal marks the position where the adsorbate film enters a resonant slice.  The two scans are taken 31 hours after each other.  The lateral offset between the scans is only $1.8\unit{nm}$, demonstrating an excellent long-term stability of the experiment.  Instrument stability and low drift are critical for recording undistorted images and avoiding artifacts in three-dimensional MRI.

To quantify the image resolution supported by our MRFM configuration we have performed $x$-scans for six resonant slices between $\frf=252.3-252.8\unit{MHz}$ (Fig.~\ref{fig:figure4}b).  By fitting the signal onset (see below) and plotting the onset position $x_0$ as a function of slice frequency $\frf$ we can directly measure the lateral field gradient at this position, $\Gx = 0.56\ee{6}\unit{T/m}$ (see SI).  Although this gradient is significantly smaller than the maximum value in Fig.~\ref{fig:figure3}a, it still allows for producing nanometer-localized signal features.

To analyze the scans, we fit the signal onset by a hyperbolic tangent step function (Fig.~\ref{fig:figure4}b, inset),
\begin{align}\label{eq:fitfunc}
\Fspin^2(x)=\Fspinmax^2 \left(1 + \mathrm{e}^{-\frac{4}{w}(x-x_0)} \right)^{-1}
\end{align}
where $x_0$ is the position of the signal onset, $w$ the characteristic width of the signal edge, and $\Fspinmax^2$ the signal step height.  We find a characteristic width for the scans in Fig.~\ref{fig:figure4}b of $w \sim 10\unit{nm}$.  This width is not indicative of the spatial resolution, however, because it is determined by the convolution of the sample with the residual curvature of the tangential imaging slice (see SI).  When the signal error is dominated by statistical fluctuations, a suitable metric for the spatial resolution must be based on the signal-to-noise ratio (SNR) of the measurement. Here, we compare the maximum signal slope to the noise of the measurement (error bars in Fig.~\ref{fig:figure4}b, inset) and obtain
\begin{align}
\sigxslope = \frac{\Fnoise^2}{\left|\partial \Fspin^2/\partial x\right|_\mathrm{max}}.
\end{align}
From a slope of $\left|\partial \Fspin^2/\partial x\right|_\mathrm{max}=14-20\unit{aN^2/nm}$ and a total noise variance of $\Fnoise^2=13-14\unit{aN^2}$, we find an uncertainty-limited spatial resolution of $\sigxslope = 0.9\pm 0.2\unit{nm}$ for a set of 11 scans.  Another method consists of computing the mean fit uncertainty $\sigxo = 0.6 \pm 0.1\unit{nm}$ of the onset positions $x_0$, from which we gain an estimate of the localization precision for these scans.  Overall, our data clearly demonstrate that MRFM is able to perform one-dimensional MRI with sub-nanometer spatial resolution.

The representative scans shown in Fig.~\ref{fig:figure4}b are taken in a tangential configuration (cf. Fig.~\ref{fig:figure3}b) to create a sharp signal feature with little convolution by the slice's point spread function.  The magnetic gradient in this configuration is, however, not very large ($\Gx\sim 0.56\ee{6}\unit{T/m}$), limiting both the sensitivity and the spatial resolution.  We can use our tip model (Fig.~\ref{fig:figure1}d,e) to estimate the maximum gradients produced in our experimental arrangement.  From the numerical model of the tip, we find a maximum $|\vecG| \approx 6\ee{6}\unit{T/m}$ at $z=10\unit{nm}$ (white star in Fig.~\ref{fig:figure3}a), with a corresponding expected slice width of $0.3\unit{nm}$.

Summarizing, we demonstrate one-dimensional scans with a resolution of 0.9~nm, a precision of 0.6~nm and a minimum measured slice width of 0.7~nm.  Our work supplies proof that the complex protocols involved in MRFM are compatible with performing MRI imaging at a sub-nanometer scale.  At the core of this improvement is a spin inversion protocol that enables sharply defined imaging slices.  We also decreased to below 1 nm all technical sources of blur in our setup, such as stage drift and sample-gradient convolution.
Our current experiment is sensitive to spin ensembles containing about $10^3-10^5$ hydrogen atoms, depending on the sample position in the gradient field. Future work will focus on improving the transducer sensitivity, such that three-dimensional images with 1 nm voxel size, corresponding to about $10^2$ hydrogen atoms, become possible.  Several routes lead towards this goal, for instance, a reduction of sensor dissipation through surface treatment and spatial design~\cite{tao14,rossi17,delepinay17,tsaturyan17,heritier18,Ghadimi18}, the use of higher magnetic field gradients~\cite{longenecker12,mamin12apl,tao16}, or a lower operating temperature~\cite{usenko11}.

\begin{acknowledgement}

We thank Ute Drechsler for her support with the clean-room fabrication.
This work was supported by Swiss National Science Foundation (SNFS) through the National Center of Competence in Research in Quantum Science and Technology (NCCR QSIT) and the Sinergia grant (CRSII5\_177198/1), and the European Research Council through the ERC Starting Grant ``NANOMRI'' (Grant agreement 309301).

\end{acknowledgement}


\providecommand{\latin}[1]{#1}
\makeatletter
\providecommand{\doi}
  {\begingroup\let\do\@makeother\dospecials
  \catcode`\{=1 \catcode`\}=2 \doi@aux}
\providecommand{\doi@aux}[1]{\endgroup\texttt{#1}}
\makeatother
\providecommand*\mcitethebibliography{\thebibliography}
\csname @ifundefined\endcsname{endmcitethebibliography}
  {\let\endmcitethebibliography\endthebibliography}{}

\end{document}